# SOLUTION FOR THE NONLINEAR TELEGRAPH EQUATION WITH TWO SPACE VARIABLES

Emad K. Jaradat[1], Read S. Hijjawi[1], Belal R. Alasassfeh[1,] Omar K. Jaradat[2]

[1] Department of Physics, Mutah University, Jordan

[2] Department of Mathematics, Mutah University, Jordan

emad_jaradat75@yahoo.com; ejaradat@mutah.edu.jo;

**Abstract:**

In this paper, we consider two space variables of nonlinear telegraph equation in terms of voltage and current. The numerical algorithm based on the Laplace transform method (LDM) is applied to obtain analytic and approximate solutions of the space-time nonlinear telegraph equations. The LDM is combined in the form of Laplace transform and Adomian Decomposition method. The results obtained by the LDM show that the approach is easy to implement and computationally very attractive.

## 1. Introduction:

Telegraph equation with two space variables is generally utilized in the investigation of wave propagation of electric signals in a cable transmission line and furthermore in wave phenomena[1]. In past years, numerous papers had talked about the arrangement of this equation by various techniques. Biazar, Ebrahimi[2] and Abdou[3] solved Famous partial differential equation using Adomian decomposition method; Sari[4] obtained approximate solution using the Daftardar-Gejji-Jafaris method (DGJ), for two-space dimensional telegraph equation, Dehghan, Ghesmati[5] used a numerical method based on the boundary integral equation (BIE) and an application of the dual reciprocity method (DRM). In this paper, Laplace decomposition method, which is combined form of the Laplace transform method and the Adomian decomposition method, used to solve the two-space dimensional nonlinear telegraph equation as:

$$\frac{\partial^2 \omega}{\partial t^2} + k\frac{\partial \omega}{\partial t} = a\frac{\partial^2 \omega}{\partial x^2} + \frac{\partial}{\partial y}\left[(b\omega + c)\frac{\partial \omega}{\partial y}\right] \qquad (1.1)$$

Rearranging this equation, we get

$$\omega_{tt} + k\omega_t = a\omega_{xx} + b\omega_y^2 + (b\omega + c)\omega_{yy} \qquad (1.2)$$

With initial conditions:

$$\omega(x,y,0) = f(x,y) \quad , \quad \omega_t(x,y,0) = g(x,y) \qquad (1.3)$$

## 2. Laplace Decomposition Method (LDM):

Consider the general form of second non-homogeneous nonlinear partial differential equations[6]

$$Lu(x,t) + Ru(x,t) + Nu(x,t) = h(x,t) \qquad (2.1)$$

Taking Laplace inverse $L^{-1}$ for both sides:

$$L^{-1}u(x,t) + L^{-1}Ru(x,t) + L^{-1}Nu(x,t) = L^{-1}h(x,t) \qquad (2.2)$$

$$L^{-1}u(x,t) = L^{-1}h(x,t) - +L^{-1}Ru(x,t) - L^{-1}Nu(x,t) \qquad (2.3)$$

The main step of the this method is to identifies the nonlinear term $Nu$ by the decomposition series[7]

$$Nu = \sum_{n=0}^{\infty} A_n(u_0, u_1, u_2, u_3, \ldots u_n) \qquad (2.4)$$

Where $A_n$ is so-called Adomian polynomials, and:

$$A_n(u_0, u_1, u_2, u_3, \ldots u_n) = \frac{1}{n!} \left[ \frac{d^n}{d\lambda^n} N\left( \sum_{i=0}^{\infty} \lambda^i u_i \right) \right] \qquad (2.5)$$

The first few Adomian polynomials given by[8]

$A_0 = N(u_0)$

$A_1 = u_1 N'(u_0)$

$A_2 = u_2 N'(u_0) + \frac{1}{2} u_1^2 N''(u_0)$

$A_3 = u_3 N'(u_0) + u_1 u_2 N''(u_0) + \frac{1}{3!} u_3 N^3(u_0)$

$A_4 = u_4 N'(u_0) + \left( \frac{1}{2} u_2^2 + u_1 u_3 \right) N''(u_0) + \frac{1}{2!} u_1^2 u_2 N^{(3)}(u_0) + \frac{1}{4!} u_1^4 N^4(u_0)$

$$(2.6)$$

## 3. Method of Solution:

In this part, we will apply the Laplace decomposition method to the two-space telegraph equation (1.2) with the conditions in (1.3).

First, we compare Eq. (1.2) with the general form in (2.1), considering that we substitute $u(x,t)$ in (2.1) by $\omega(x,y,t)$, we find that:

$$L\omega(x,y,t) = h(x,t) - R\omega(x,y,t) - N\omega(x,y,t) \qquad (3.1)$$

With:

$$L = \frac{\partial^2}{\partial t^2}, R = a\frac{\partial^2}{\partial x^2} \text{ And } h(x,t) = 0 \qquad (3.2)$$

While the nonlinear term given by:

$$N\omega = b\omega_y^2 + (b\omega + c)\omega_{yy} \qquad (3.3)$$

Applying Laplace transform on both sides of Eq (1.2) with respect to $t$ variable, we have:

$$s^2\omega(x,y,s) - sf(x,y) - g(x,y) + k\,s\omega(x,y,s) - k\,f(x,y)$$
$$= \mathcal{L}\{a\,\omega_{xx} + b\,\omega_y^2 + (b\omega + c)\omega_{yy}\}$$

$$(s^2 + k\,s)\,\omega(x,y,s) = (s+k)f(x,y) + g(x,y) + \mathcal{L}\{a\,\omega_{xx} + b\,\omega_y^2 + (b\omega + c)\omega_{yy}\}$$

$$\omega(x,y,s) = \frac{1}{s^2 + k\,s}[(s+k)f(x,y) + g(x,y) + \mathcal{L}\{a\,\omega_{xx} + b\,\omega_y^2 + (b\omega + c)\omega_{yy}\}]$$

$$\omega(x,y,s) = \frac{1}{s}f(x,ys) + \frac{1}{s^2 + k\,s}g(x,y,s)$$
$$+ \frac{1}{s^2 + k\,s}\mathcal{L}\{a\,\omega_{xx} + b\,\omega_y^2 + (b\omega + c)\omega_{yy}\} \qquad (3.4)$$

Taking inverse Laplace transformation to both sides of Eq. (3.4) with respect to $t$ variable we have:

$$\omega(x,y,t) = f(x,y,t) + \frac{1 - e^{-kt}}{k}g(x,y,t)$$
$$+ \mathcal{L}^{-1}\left\{\frac{1}{s^2 + k\,s}\mathcal{L}\{a\,\omega_{xx} + b\,\omega_y^2 + (b\omega + c)\omega_{yy}\}\right\} \qquad (3.5)$$

The Laplace decomposition method (LDM)[7] assumes a series solution of the function u(x, t) given by:

$$u(x,t) = \sum_{n=0}^{\infty} u_n(x,t) \qquad (3.6)$$

For our equation, we get:

$$\omega(x,y,t) = \sum_{n=0}^{\infty} \omega_n(x,y,t) \qquad (3.7)$$

Substituting Eq. (2.4) and Eq. (3.7) in Eq. (3.5), we get:

$$\sum_{n=0}^{\infty} \omega_n(x,y,t) = f(x,y,t) + \frac{1 - e^{-kt}}{k}g(x,y,t)$$
$$+ \mathcal{L}^{-1}\left\{\frac{1}{s^2 + k\,s}\mathcal{L}\left\{\sum_{n=0}^{\infty} a\,\omega_{nxx} + \sum_{n=0}^{\infty} A_n\right\}\right\} \qquad (3.8)$$

Comparing the both sides of the previous equation, we get:

$$\omega_0(x,y,t) = f(x,y,t) + \frac{1-e^{-kt}}{k}g(x,y,t) \tag{3.9}$$

$$\omega_1(x,y,t) = \mathcal{L}^{-1}\left\{\frac{1}{s^2+ks}\mathcal{L}\left\{\sum_{n=0}^{\infty} a\,\omega_{0xx} + \sum_{n=0}^{\infty} A_0\right\}\right\}$$

$$\omega_2(x,y,t) = \mathcal{L}^{-1}\left\{\frac{1}{s^2+ks}\mathcal{L}\left\{\sum_{n=0}^{\infty} a\,\omega_{1xx} + \sum_{n=0}^{\infty} A_1\right\}\right\}$$

$$\omega_{n+1}(x,y,t) = \mathcal{L}^{-1}\left\{\frac{1}{s^2+ks}\mathcal{L}\left\{\sum_{n=0}^{\infty} a\,\omega_{nxx} + \sum_{n=0}^{\infty} A_n\right\}\right\} \tag{3.10}$$

Now note that the nonlinear term in (3.3), $N\omega = b\omega^2{}_y + (b\omega + c)\omega_{yy}$ can be split into two terms which making the calculation simpler

$N_1 = b\omega_y^2$

$N_2 = (b\omega + c)\omega_{yy}$

From this, we will now consider the decomposition of the nonlinear terms into Adomian polynomials s:

$$N_1\omega = b\omega_y^2 = \sum_{n=0}^{\infty} P_n(u_0, u_1, u_2, \ldots, u_n) \tag{3.11}$$

$$N_2\omega = (b\omega + c)\omega_{yy} = \sum_{n=0}^{\infty} Q_n(u_0, u_1, u_2, \ldots, u_n) \tag{3.12}$$

Using Eq. (3.6) of ADM, we get:

We can calculate that:

$N_1(\omega) = b\omega_y^2$

$$= b\big(\omega_{0y} + \omega_{1y} + \omega_{2y} + \omega_{3y} + \omega_{4y} + \cdots\big)^2$$

$$= b\,(\omega_{0y}^2 + 2\omega_{0y}\omega_{1y} + 2\omega_{0y}\omega_{2y} + 2\omega_{0y}\omega_{3y} + 2\omega_{0y}\omega_{4y} + \omega_{1y}^2 + 2\omega_{1y}\omega_{2y} + 2\omega_{1y}\omega_{3y} + 2\omega_{1y}\omega_{4y} + \omega_{2y}^2 + 2\omega_{2y}\omega_{3y} + 2\omega_{2y}\omega_{4y} + \omega_{3y}^2 + 2\omega_{3y}\omega_{4y} + \omega_{4y}^2) \tag{3.13}$$

The above expressions can be rearrange by grouping terms in which the sum of subscripts of $\omega_n$ be the same. This procedure gives the Adomian polynomials for $N_1$ and $N_2$, now for $N_1$ we get[9-11]:

$P_0 = b\omega_{0y}^2$

$P_1 = 2b\omega_{0y}\omega_{1y}$

$P_2 = b\omega_{1y}^2 + 2b\omega_{0y}\omega_{2y}$

$P_3 = 2b\omega_{0y}\omega_{3y} + 2b\omega_{1y}\omega_{2y}$

$P_4 = b\omega_{2y}^2 + 2b\omega_{0y}\omega_{4y} + 2b\omega_{1y}\omega_{3y}$

Now by the same way we did to find Adomian polynomials for $N_1$, we will find the polynomials for $N_2$:

$N_2\omega = (b\omega + c)\omega_{yy}$

$= \big((b\omega_0 + c) + (b\omega_1 + c) + (b\omega_2 + c) + (b\omega_3 + c) + (b\omega_4 + c)\big)$

$\times \big(\omega_{0y} + \omega_{1y} + \omega_{2y} + \omega_{3y} + \omega_{4y}\big)$

$= (b\omega_0 + c)\omega_{0y} + (b\omega_0 + c)\omega_{1y} + (b\omega_0 + c)\omega_{2y} + (b\omega_0 + c)\omega_{3y} + (b\omega_0 + c)\omega_{4y} + (b\omega_1 + c)\omega_{0y} + (b\omega_1 + c)\omega_{1y} + (b\omega_1 + c)\omega_{2y} + (b\omega_1 + c)\omega_{3y} + (b\omega_1 + c)\omega_{4y} + (b\omega_2 + c)\omega_{0y} + (b\omega_2 + c)\omega_{1y} + (b\omega_2 + c)\omega_{2y} + (b\omega_2 + c)\omega_{3y} + (b\omega_2 + c)\omega_{4y} + (b\omega_3 + c)\omega_{0y} + (b\omega_3 + c)\omega_{1y} + (b\omega_3 + c)\omega_{2y} + (b\omega_3 + c)\omega_{3y} + (b\omega_3 + c)\omega_{4y} + (b\omega_4 + c)\omega_{0y} + (b\omega_4 + c)\omega_{1y} + (b\omega_4 + c)\omega_{2y} + (b\omega_4 + c)\omega_{3y} + (b\omega_4 + c)\omega_{4y}$ (3.14)

Now we can evaluate the Adomian polynomials for $N_2$:

$Q_0 = (b\omega_0 + c)\omega_{0y}$

$Q_1 = (b\omega_0 + c)\omega_{1y} + (b\omega_1 + c)\omega_{0y}$

$Q_2 = (b\omega_0 + c)\omega_{2y} + (b\omega_1 + c)\omega_{1y} + (b\omega_2 + c)\omega_{0y}$

$Q_3 = (b\omega_0 + c)\omega_{3y} + (b\omega_1 + c)\omega_{2y} + (b\omega_2 + c)\omega_{1y} + (b\omega_3 + c)\omega_{0y}$

$Q_4 = (b\omega_0 + c)\omega_{4y} + (b\omega_1 + c)\omega_{3y} + (b\omega_2 + c)\omega_{2y} + (b\omega_3 + c)\omega_{1y} + (b\omega_4 + c)\omega_{0y}$

Now considering Eq. (3.11) and Eq. (3.12), we have:

$$Nu = \sum_{n=0}^{\infty} A_n(u_0, u_1, u_2, \ldots u_n) = \sum_{n=0}^{\infty} \big((P_n + Q_n)(u_0, u_1, u_2, \ldots u_n)\big) \quad (3.15)$$

Therefore, the Adomian polynomials are:

$$A_0 = b\omega_{0y}^2 + (b\omega_0 + c)\omega_{0y}$$

$$A_1 = 2b\omega_{0y}\omega_{1y} + (b\omega_0 + c)\omega_{1y} + (b\omega_1 + c)\omega_{0y}$$

$$A_2 = b\omega_{1y}^2 + 2b\omega_{0y}\omega_{2y} + (b\omega_0 + c)\omega_{2y} + (b\omega_1 + c)\omega_{1y} + (b\omega_2 + c)\omega_{0y}$$

$$A_3 = 2b\omega_{0y}\omega_{3y} + 2b\omega_{1y}\omega_{2y} + (b\omega_0 + c)\omega_{3y} + (b\omega_1 + c) + (b\omega_2 + c)\omega_{1y} + (b\omega_3 + c)\omega_{0y}$$

$$A_4 = b\omega_{2y}^2 + 2b\omega_{0y}\omega_{4y} + 2b\omega_{1y}\omega_{3y} + (b\omega_0 + c)\omega_{4y} + (b\omega_1 + c)\omega_{3y} + (b\omega_2 + c)\omega_{2y} + (b\omega_3 + c)\omega_{1y} + (b\omega_4 + c)\omega_{0y}$$

### 4. Applications:

In this section, we will solve the two-space telegraph equation in Eq. (1.2) by taking the following initial conditions:

1) $\omega(x, y, 0) = f(x, y) = \sin x + \sin y$

$\omega_t(x, y, 0) = g(x, y) = 0$

$a = b = 1, c = 0$

From Eq. (3.9), we get:

$\omega_0 = \sin x + \sin y$

From relation in (3.10), to find $\omega_1$ we need $A_0$:

$$A_0 = \omega_{0y}^2 + \omega_0\omega_{0y}$$

$$= \cos^2 y + b(\sin x + \sin y)\cos y$$

$$= \cos^2 y + \sin x \cos y + \sin y \cos y$$

$$\omega_1 = \mathcal{L}^{-1}\left\{\frac{1}{s^2 + k\,s}\mathcal{L}\{a\,\omega_{0xx} + A_0\}\right\}$$

$$= \mathcal{L}^{-1}\left\{\frac{1}{s^2 + k\,s}\mathcal{L}\{-\sin x + \cos^2 y + \sin x \cos y + \sin y \cos y\}\right\}$$

$$= \mathcal{L}^{-1}\left\{\frac{1}{s^3 + k\,s^2}\{-\sin x + \cos^2 y + \sin x \cos y + \sin y \cos y\}\right\}$$

$$= \frac{e^{-kt} + kt - 1}{k^2}\{-\sin x + \cos^2 y + \sin x \cos y + \sin y \cos y\}$$

$$A_1 = 2\omega_{0y}\omega_{1y} + \omega_0\omega_{1y} + \omega_1\omega_{0y}$$

$$= \frac{e^{-kt} + kt - 1}{k^2}(2\cos y(-2\cos y \sin y - \sin x \sin y + \cos^2 y - \sin^2 y)$$
$$+ (\sin x + \sin y)(-2\cos y \sin y - \sin x \sin y + \cos^2 y - \sin^2 y)$$
$$+ \cos y(-\sin x + \cos^2 y + \sin x \cos y + \sin y \cos y))$$

$$\omega_2 = \mathcal{L}^{-1}\left\{\frac{1}{s^2 + k\,s}\mathcal{L}\{a\,\omega_{1xx} + A_1\}\right\}$$

$$= \mathcal{L}^{-1}\left\{\frac{1}{s^2 + k\,s}\mathcal{L}\left\{\frac{e^{-kt} + kt - 1}{k^2}(-\cos x + \cos x \cos y)\right.\right.$$
$$+ \frac{e^{-kt} + kt - 1}{k^2}(2\cos y(-2\cos y \sin y - \sin x \sin y + \cos^2 y - \sin^2 y)$$
$$+ (\sin x + \sin y)(-2\cos y \sin y - \sin x \sin y + \cos^2 y - \sin^2 y)$$
$$\left.\left.+ \cos y(-\sin x + \cos^2 y + \sin x \cos y + \sin y \cos y))\right\}\right\}$$

$$\omega_2 = \mathcal{L}^{-1}\left\{\frac{1}{s^2 + k\,s}\mathcal{L}\left\{\frac{e^{-kt} + kt - 1}{k^2}((-\cos x + \cos x \cos y)\right.\right.$$
$$+ (2\cos y(-2\cos y \sin y - \sin x \sin y + \cos^2 y - \sin^2 y)$$
$$+ (\sin x + \sin y)(-2\cos y \sin y - \sin x \sin y + \cos^2 y - \sin^2 y)$$
$$\left.\left.+ \cos y(-\sin x + \cos^2 y + \sin x \cos y + \sin y \cos y))\right\}\right\}$$

$$= \mathcal{L}^{-1}\left\{\frac{1}{s^2 + k\,s} * \frac{1}{s^3 + k\,s^2}((-\cos x + \cos x \cos y)\right.$$
$$+ (2\cos y(-2\cos y \sin y - \sin x \sin y + \cos^2 y - \sin^2 y)$$
$$+ (\sin x + \sin y)(-2\cos y \sin y - \sin x \sin y + \cos^2 y - \sin^2 y)$$
$$\left.+ \cos y(-\sin x + \cos^2 y + \sin x \cos y + \sin y \cos y))\right\}$$

$$= \frac{6 + kt(-4 + kt) - 2e^{-kt}(3 + kt)}{2k^4}((-\cos x + \cos x \cos y)$$
$$+ (2\cos y(-2\cos y \sin y - \sin x \sin y + \cos^2 y - \sin^2 y)$$
$$+ (\sin x + \sin y)(-2\cos y \sin y - \sin x \sin y + \cos^2 y - \sin^2 y)$$
$$+ \cos y(-\sin x + \cos^2 y + \sin x \cos y + \sin y \cos y))$$

Thus, the approximate of the two space nonlinear telegraph equation (1.1) given by:

$$\omega_{LDM} = \omega_0(x, y, t) + \omega_1(x, y, t) + \omega_2(x, y, t) + \cdots \qquad (4.1)$$

2) $\omega(x, y, 0) = f(x, y) = e^x + e^y$

$\omega_t(x, y, 0) = g(x, y) = 0$

$a = b = 1, c = 0$

From Eq. (3.9), we get:

$\omega_0 = e^x + e^y$

Now we find $A_0$ to find $\omega_1$:

$A_0 = \omega_{0y}^2 + \omega_0 \omega_{0y}$

$A_0 = 2 e^{2y} + e^{x+y}$

So:

$\omega_1 = \mathcal{L}^{-1}\left\{\frac{1}{s^2 + k s} \mathcal{L}\{a\, \omega_{0xx} + A_0\}\right\}$

$= \mathcal{L}^{-1}\left\{\frac{1}{s^3 + k s^2}\{e^x + 2 e^{2y} + e^{x+y}\}\right\}$

$= \frac{e^{-kt} + kt - 1}{k^2}\{e^x + 2 e^{2y} + e^{x+y}\}$

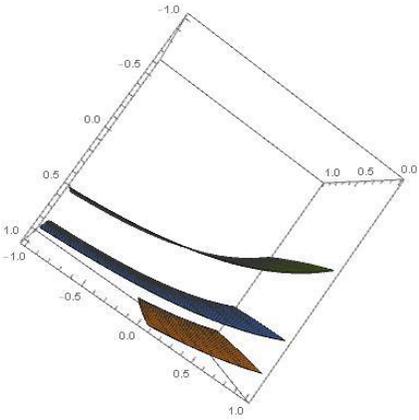

Fig. 1. Space-time graph of the solution term $\omega_1$ of telegraph equation with $k = 1$

Now to find $\omega_2$, first to evaluate $A_1$:

$A_1 = 2\omega_{0y}\omega_{1y} + \omega_0 \omega_{1y} + \omega_1 \omega_{0y}$

$= \frac{e^{-kt} + kt - 1}{k^2}(2 e^y(4 e^{2y} + e^{x+y}) + (e^x + e^y)(4 e^{2y} + e^{x+y})$
$\quad + e^y(e^x + 2 e^{2y} + e^{x+y}))$

$$= \frac{e^{-kt} + kt - 1}{k^2} (14\, e^{3y} + 8\, e^{x+2y} + e^{2x+y} + e^{x+y})$$

Now we can evaluate $\omega_2$:

$$\omega_2 = \mathcal{L}^{-1}\left\{\frac{1}{s^2 + k\,s} \mathcal{L}\{a\, \omega_{1xx} + A_1\}\right\}$$

$$= \mathcal{L}^{-1}\left\{\frac{1}{s^2 + k\,s} \mathcal{L}\left\{\frac{e^{-kt} + kt - 1}{k^2}(e^x + e^{x+y})\right.\right.$$

$$\left.\left. + \frac{e^{-kt} + kt - 1}{k^2}(14\, e^{3y} + 8\, e^{x+2y} + e^{2x+y} + e^{x+y})\right\}\right\}$$

$$= \mathcal{L}^{-1}\left\{\frac{1}{s^2 + k\,s} \mathcal{L}\left\{\frac{e^{-kt} + kt - 1}{k^2}(e^x + 2\, e^{x+y} + 14\, e^{3y} + 8\, e^{x+2y} + e^{2x+y})\right\}\right\}$$

$$= \mathcal{L}^{-1}\left\{\frac{1}{s^2 + k\,s} * \frac{1}{s^3 + k\,s^2}(e^x + 2\, e^{x+y} + 14\, e^{3y} + 8\, e^{x+2y} + e^{2x+y})\right\}$$

$$= \frac{6 + kt(-4 + kt) - 2e^{-kt}(3 + kt)}{2k^4}(e^x + 2\, e^{x+y} + 14\, e^{3y} + 8\, e^{x+2y} + e^{2x+y})$$

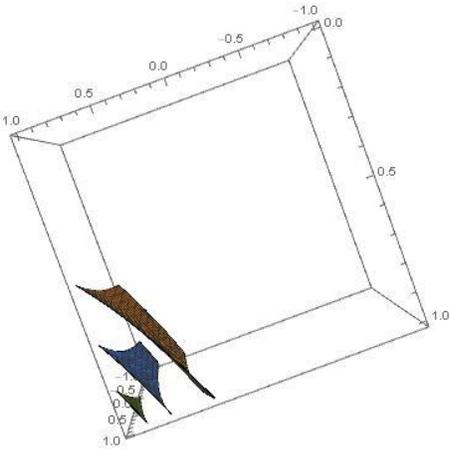

Fig. 2. Space-time graph of the solution term $\omega_2$ of telegraph equation with $k = 1$

We can find the approximate of the two-space nonlinear telegraph equation for this case also by (4.1):

$$\omega_{LDM} = \omega_0(x, y, t) + \omega_1(x, y, t) + \omega_2(x, y, t) + \cdots$$

It can be seen from the Figures that the solution obtained by the present method is nearly identical with the exact solution. It is to be noted that only the second order term of the Laplace Decomposition Method is used in evaluating the approximate solutions for Figures 1, 2. It is evident that the efficiency of this approach can be dramatically enhanced by computing further terms of $w(x,t)$ when the LDM is used.

**5. Conclusion:**

In this work, the LDM has been successfully applied for solving two dimensional nonlinear telegraph equations with boundary conditions. This method of solution provides the solutions in terms of a convergent series with easily calculable components in a direct way without using perturbation. It is worth mentioning that the LDM is capable of reducing the volume of the computational work as compared to classical methods .Hence, we conclude that the LDM is very powerful and efficient in finding analytical and approximate solutions as well as numerical solutions for wide classes of partial differential equations.

**References:**

[1] Hind Al-badrani , Sharefah Saleh , H. O. Bakodah and M. Al-Mazmumy, Numerical Solution for Nonlinear Telegraph Equation by Modified Adomian Decomposition Method, Nonlinear Analysis .and Differential Equations, Vol. 4, 2016, no. 5, 243 – 257

[2] J. Biazar and H. Ebrahimi, An Approximation to the Solution of Telegraph Equation by Adomian Decomposition Method, International Mathematical Forum, 2 (2007), no. 45, 2231-2236.

[3] M.A. Abdou, Adomian decomposition method for solving the telegraph equation in charged particle transport, Journal of Quantitative Spectroscopy & Radiative Transfer 95 (2005) 407–414

[4] Murat Sari , Abdurrahim Gunay and Gurhan Gurarslan, A Solution to the Telegraph Equation by Using DGJ Method, International Journal of Nonlinear Science    Vol.17(2014) No.1,pp.57-66

[5 M. Dehghan and A. Ghesmati, Solution of the second-order one-dimensional hyperbolic telegraph equation by using the dual reciprocity boundary integral equation (DRBIE) method, Engineering Analysis with Boundary Elements, 34 (2010), 51-59

[6] M.Hussain  and Majid Khan, Modified Laplace Decomposition Method, Applied Mathematical Science, Vol. 4, 2010, no. 36, 1769-1783

[7] Wazwaz, A.W., Anew technique for calculating Adomian polynomials for nonlinear polynomials. Appl. Math. Comput., 111: 33-51 (2002)

[8] G. Adomian. Review of the Decomposition Method in Applied Mathematics. JOURNAL OF MATHEMATICAL ANALYSIS AND APPLICATIONS 135, 501-544 (1988)

[9] O. González-Gaxiola, J. Ruiz de Chávez and R. Bernal-Jaquez, Solution of the Nonlinear Kompaneets Equation through the Laplace-Adomian Decomposition Method. International Journal of Applied and Computational Mathematics, vol 3, 2017, 489-504.

[10] Manzoor Ahmad, Altaf Ahmad Bhat and Renu Jain, Space Time Fractional Telegraph Equation and its Application by Using Adomian Decomposition Method. no 22, 2018, 73-81.

[11] Emad K. Jaradat, Amer D. Aloqali and Wajd Alhabashneh, Using Laplace Decomposition Method To Solve Nonlinear Klien-Gordon Equation. UPB Scientific Bulletin, Series D: Mechanical Engineering, vol 180, iss 2, 2018.